\documentclass[referee]{raa}
\usepackage{graphicx,times}             
\usepackage{natbib}

\PassOptionsToPackage{normalem}{ulem}
\usepackage{ulem}
\usepackage{amssymb,amsmath}
\usepackage{multirow}
\usepackage{booktabs}
\usepackage{color}
\usepackage{ulem}
\usepackage{threeparttable}
\bibpunct{(}{)}{;}{a}{}{,}
\usepackage{xcolor}

\usepackage[colorlinks=true,linkcolor=green,anchorcolor=red,citecolor=blue,filecolor=red,urlcolor=red]{hyperref}
\begin{document}
   \title{An analysis of the fragmentation of observing time at the Muztagh-ata site
}

   \volnopage{Vol.0 (20xx) No.0, 000--000}      
   \setcounter{page}{1}          

   \author{Wen-bo Gu
      \inst{1,2}
   \and Jing Xu$*$
      \inst{1}
	\and Guo-jie Feng
		\inst{1}
	\and Xuan Zhang
		\inst{1}
   \and Le-tian Wang
		\inst{1}
	\and Xin-liang Wang
		\inst{1}
	\and Ali Esamdin
		\inst{1}
    \and li-xian Shen
		\inst{1,2}
}

   \institute{Xinjiang Astronomical Observatory, Chinese Academy of Sciences, 
				Urumqi, 830011, China; {\it guwenbo@xao.ac.cn, xujing@xao.ac.cn}\\
 		\and
             School of Astronomy and Space Science, University of Chinese Academy of Sciences, Beijing 100049, People's Republic of China\\
\vs\no
   {\small Received~~20xx month day; accepted~~20xx~~month day}}
\abstract{Cloud cover plays a pivotal role in assessing observational conditions for astronomical site-testing.  Except for the fraction of observing time, its fragmentation also wields a significant influence on the quality of nighttime sky clarity. In this article, we introduce the function $\Gamma\in \left [ 0,1 \right ]$, designed to comprehensively capture both the fraction of available observing time and its continuity. Leveraging in situ measurement data gathered at the Muztagh-ata site between 2017 and 2021, we showcase the effectiveness of our approach. The statistical result illustrates that the Muztagh-ata site affords approximately 122 nights of absolute clear and 205 very good nights annually, corresponding to $\Gamma\geq 0.9$ and $\Gamma\geq  0.36$  respectively.
}

   \authorrunning{Gu et al}            

   \maketitle

\section{Introduction}           
\label{sect:intro}
In the pursuit of identifying the most appropriate astronomical site to host large optical telescopes, a comprehensive evaluation of observing conditions becomes imperative. Factors ranging from cloud coverage and seeing to meteorological parameters necessitate thorough consideration. Among these, cloud coverage emerges as the paramount parameter when assessing potential sites\citep{Tapia1992,Graham2005,Sarazin2006}. This parameter directly determines the available observing time (hereafter referred to as AOT) of telescopes. The enduring practice of cloud coverage monitoring runs through the procedure of site-testing for the prominent grand telescope projects\citep{2009PASP..121..384S,2012MNRAS.420.1273C,2014PASP..126..412V,2020RAA....20...81C}.

A multitude of techniques are employed in the monitoring of cloud coverage to quantify and assess the fraction of AOT within site-testing endeavors. One approach utilizing data extracted from satellite-borne instruments focuses on identifying the AOT in multiband image pixels. \cite{2011MNRAS.411.1271C} presented a new homogeneous methodology to quantify the observing nights for several potential sites in the western hemisphere by the data extracted from GOES12. \cite{2022MNRAS.511.5363W} devised a method to recognize observing nights and analyze for astronomical sites in Indonesia and China mainly rooted in the FY-2 satellites. Notably, the augmentation of satellite data in terms of both time and spatial resolution has propelled more precise identification of observing night\citep{2015ExA....39..547A}. However, the availability of high-resolution satellite data coverage is not universal, On-site measurements are still a valuable supplement. \citep{2016PASP..128j5003T,2020RAA....20..149X,2021Natur.596..353D}. Sequence image data captured by all-sky cameras, subjected to manual or automated processing, are used to analyze the fraction of AOT\citep{2008SPIE.7012E..24S}. Irrespective of the data source employed, these methods evaluate the AOT at the different temporal resolutions, without accounting for the fragmentation of observing time periods.

In the earlier paper, we reported on the statistical outcomes of the cloud coverage during the period from 2017 to 2021, as derived from all-sky cameras situated at the Muztagh-ata site. With reference to the definitions of clear and cloudy nights posited in works such as \cite{1973PASP...85..255M} and \cite{Tapia1992}, we calculated the fraction of AOT and observing nights for our site. 
The Muztagh-ata site can provide 227 observing nights, 175 clear nights and 169 nights with continuous observations lasting over 4 hours annually\citep{2023RAA....23d5015X}.
From an astronomical perspective, we assert that continuous observing time (hereafter referred to as COT) holds as much significance as the fraction of AOT for an observatory. In this manuscript, we present a fresh criterion that takes into account both the fractions of AOT and the degree of their fragmentation, thereby enabling a quantitative analysis of the quality of nighttime sky clarity at the Muztagh-ata site. The organization of this paper unfolds as follows: Section 2 outlines our methodology and presents the corresponding validation. Statistical results of this method for our site are given in Section 3. The conclusions are summarized in Section 4.

\begin{figure*}[h]
   \centering
  \includegraphics[height=10cm, angle=0]{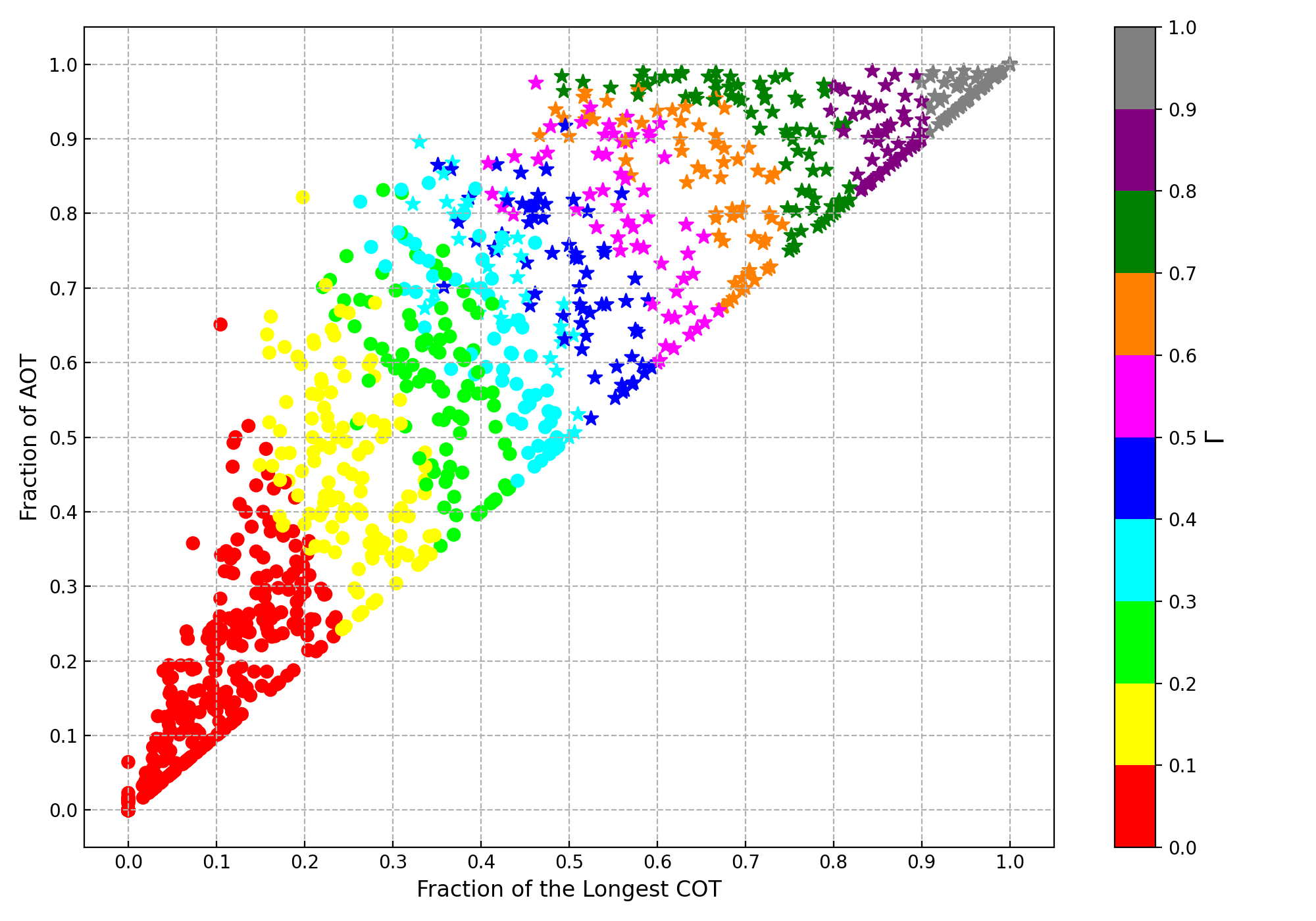}
   \caption{The distribution of Fraction of AOT and Fraction of the longest COT for a total of 1598 nights from 2017 to 2021 is illustrated. The colors represent different ranges of Gamma (\(\Gamma\)) values: red for \(\Gamma \in [0,1)\), green for \(\Gamma \in [2,3)\), cyan for \(\Gamma \in [3,4)\), blue for \(\Gamma \in [4,5)\), pink for \(\Gamma \in [5,6)\), orange for \(\Gamma \in [6,7)\), dark green for \(\Gamma \in [7,8)\), purple for \(\Gamma \in [8,9)\), and gray for \(\Gamma \in [9,10)\).The color bar in the graph also displays the color annotations corresponding to different \(\Gamma\) ranges. We have marked the nights where \(\Gamma \geq 0.36\) with asterisks (\(\ast\)) and the nights where \(\Gamma < 0.36\) with dots (\(\cdot\)) for distinction.}
   \label{fig:f1}
\end{figure*}

\begin{figure*}[h]
   \centering
  \includegraphics[height=15cm, angle=0]{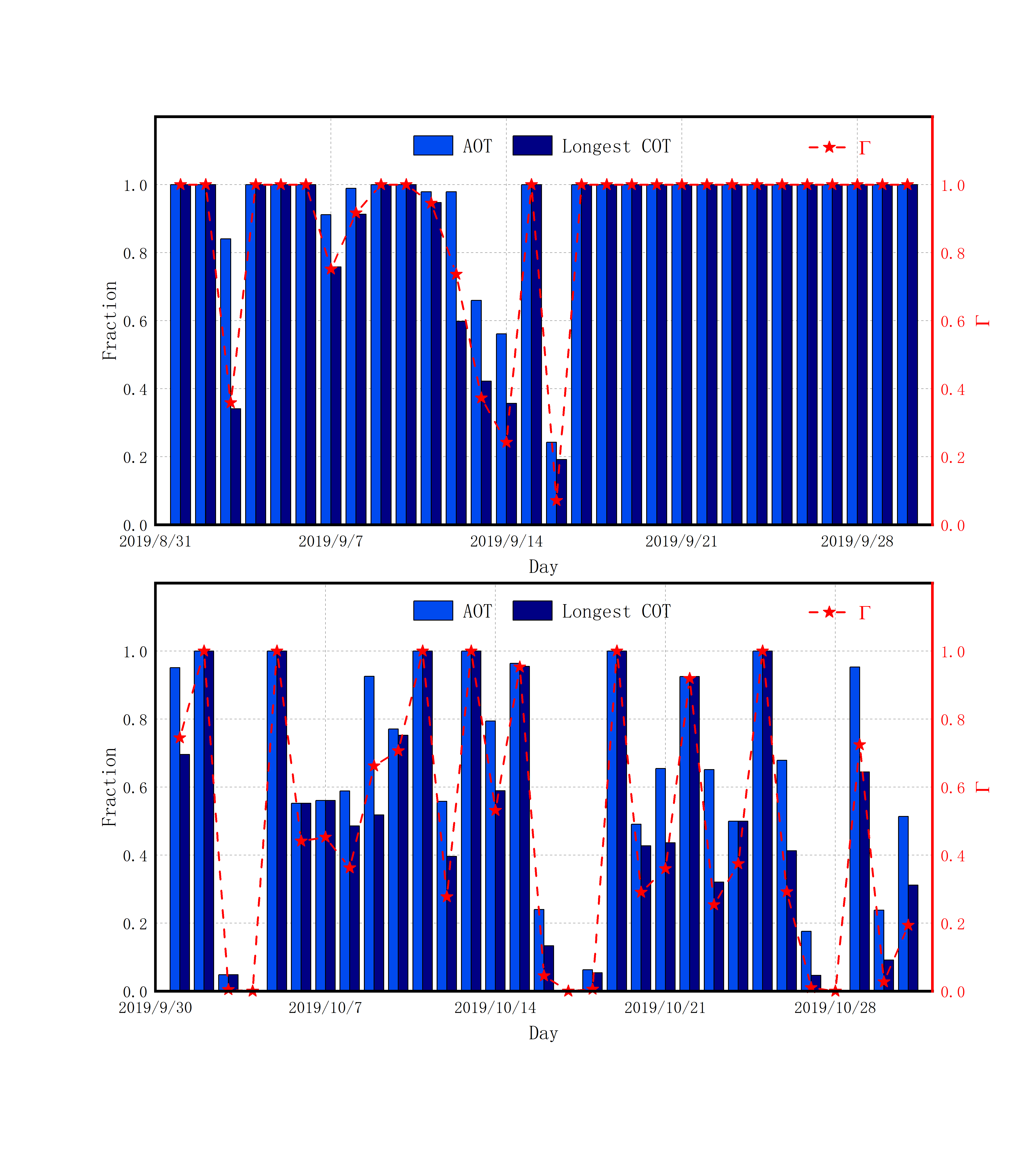}
   \caption{$\Gamma$ (red stars) and the fraction of AOT (light blue) and the longest COT (dark blue) for the nights in 2019 September and October.}
   \label{fig:f2}
\end{figure*}

\section{Method}
The presence of drifting clouds in the night sky above astronomical observatories can significantly diminish the quality and effectiveness of telescope observations. This effect becomes particularly pronounced in certain types of observations, such as time-domain astronomy, where the intermittent passage of clouds leads to the loss of valuable information. Even among nights with comparable fractions of AOT, there are variations in the extent of COT. It is unequivocal that the most optimal night is the one characterized by the longest COT. Thus, it becomes imperative to evaluate the quality of nights by considering both the fraction of AOT and the degree of its fragmentation. To address this concern, we introduce a criterion function formulated as follows:

\begin{equation}
\label{eq:e1}
\Gamma =\frac{\sum k_{i}\omega _{i}  }{N_{T} },\omega _{i}=-\left ( \frac{k_{i}}{N_{T}} \right ) ^{2}+2\left ( \frac{k_{i}}{N_{T}} \right )
\end{equation}
\begin{figure*}[h]
   \centering
  \includegraphics[height=2.2cm, angle=0]{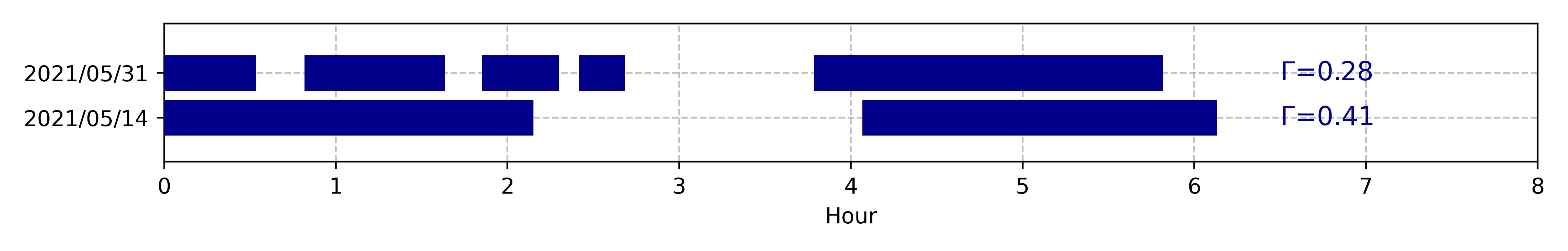}
   \caption{The hourly distribution of observable time on the nights of May 13th and May 30th, 2021, is depicted. The blue areas represent the observable time, with 2 segments of observability on May 14th and 5 segments on May 30th. The corresponding $\Gamma$ values for each night are indicated in the graph.}
   \label{fig:yun}
\end{figure*}

where $k_{i}$ is the length of the $i\textsuperscript{th}$ COT of one night, $\omega _{i}\in \left [ 0,1 \right ] $ is the weight, and $N_{T}$ is the length of the night. $\Gamma$ can more accurately describe the quality of sky clarity during the whole night than only using the length of AOT or COT. 
Our dataset is sampled at 5-minute intervals. To better define continuous observation time periods, we have established a threshold: If observations are interrupted for five minutes or longer due to cloud cover within a time interval, we consider our continuous observations to be interrupted. In addition, if a brief observable time occurs during a continuous Non-observable time, we consider there must be at least 10 minutes or more of observable time to acknowledge the presence of an observable time interval.
During the period from 2017 to 2021, We calculate the value of $\Gamma$ for the nights all had a record of all-sky images for more than 50\% of the nighttime. Finally we obtained statistically significant data for 1598 $\Gamma$ functions.
The scatterplot in Figure~\ref{fig:f1}  displays the fraction of AOT values and the longest COT duration for each of the 1598 nights we calculated, corresponding to different $\Gamma$ values.
In Figure ~\ref{fig:f1}, the heatmap uses different colors to represent $\Gamma$ values in distinct value ranges. To enhance visibility, we have marked data points with $\Gamma\geq 0.36$ using asterisks (*) and data points with $\Gamma < 0.36$ using dots (·).
We use \(\Gamma = 0.36\) as a threshold, on nights with \(\Gamma \geq 0.36\), the site can offer at least 50\% AOT and continuous observations for over 32\% of the night , averaging above 60\% AOT and over 40\% continuous observation time. This also simultaneously meets the criteria we previously defined for observing nights\citep{2023RAA....23d5015X}. 
Therefore, we define nights with \(\Gamma \geq 0.36\) as very good nights with relatively high observability and observational continuity and those with \(\Gamma\geq 0.9\) as absolute clear nights, during which observations can essentially be carried out throughout the night.

The $\Gamma$ of each night during September and October 2019 are exhibited in Figure~\ref{fig:f2}. This figure includes the fractions of AOT and the duration of the longest COT for each night. A noteworthy instance is observed from September 17 to 30, where continuous clear conditions prevailed throughout the entire night, yielding $\Gamma=1$. The $\Gamma$ value for October 8 was lower compared to October 6. This discrepancy primarily arises due to the fragmentation of AOT during the former night, despite having a slightly higher fraction of AOT than the latter. 
Figure~\ref{fig:yun} illustrates the AOT distribution on the nights of May 13th and May 30th, 2021. These two nights had nearly identical AOT and the longest COT. However, due to more fragmented observable periods on the night of May 30th, the $\Gamma$ value was 0.28. The same AOT and the longest COT can result in different $\Gamma$ values due to varying degrees of fragmentation. It is worth mentioning that due to the geographical location of the observatory site, astronomical twilight occurs relatively late in the summer, sometimes approaching midnight. That is also why the start times for the two nights in Figure~\ref{fig:yun} are both at 0:00 of the following day.
Both Figure~\ref{fig:f1}, Figure~\ref{fig:f2} and Figure~\ref{fig:yun} serve to substantiate the practical efficacy of this method.

\begin{figure*}[h]
   \centering
  \includegraphics[height=6cm, angle=0]{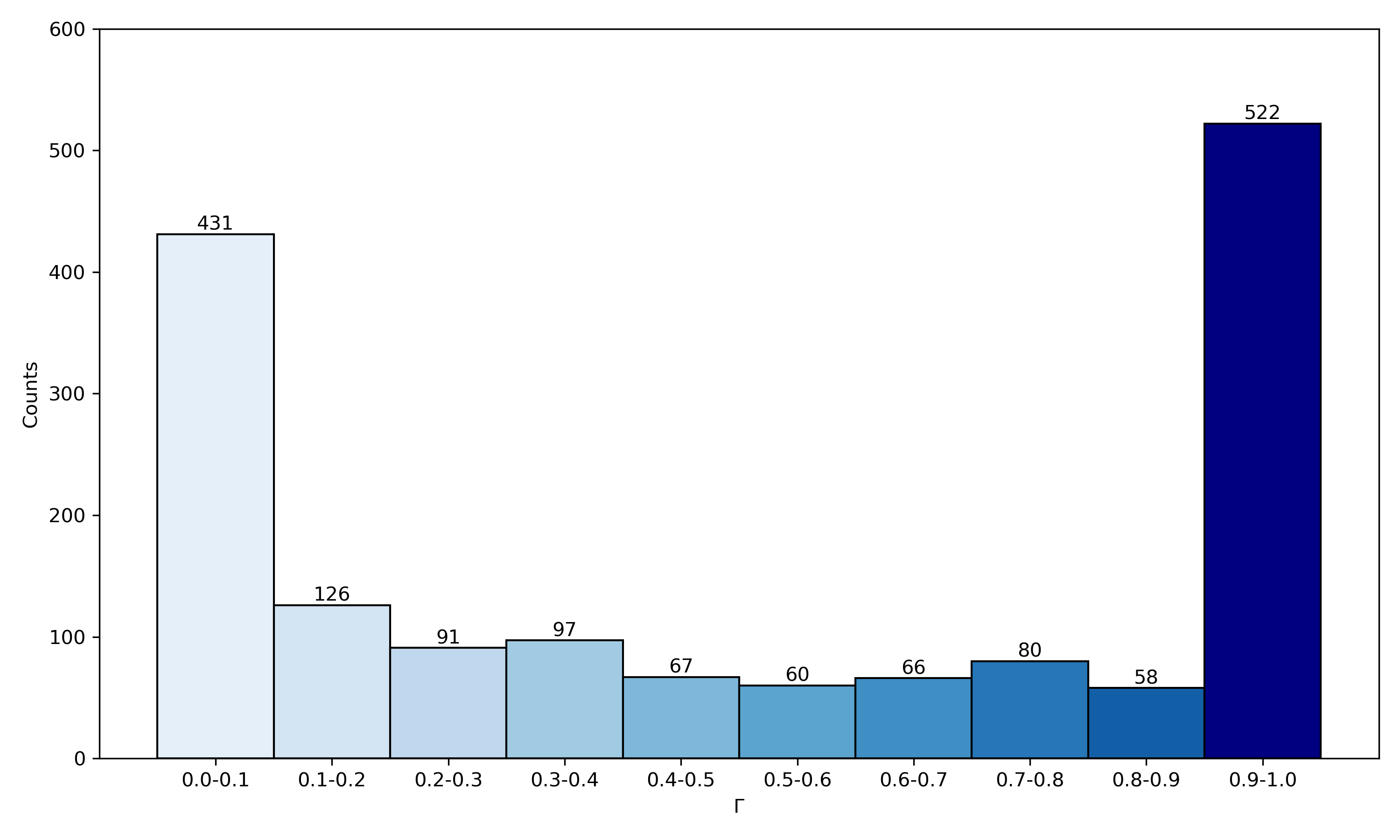}
   \caption{Statistical of $\Gamma$ for the 1598 nights from 2017 to 2021.}
   \label{fig:f3}
\end{figure*}
\begin{figure*}[h]
   \centering
  \includegraphics[height=8cm, angle=0]{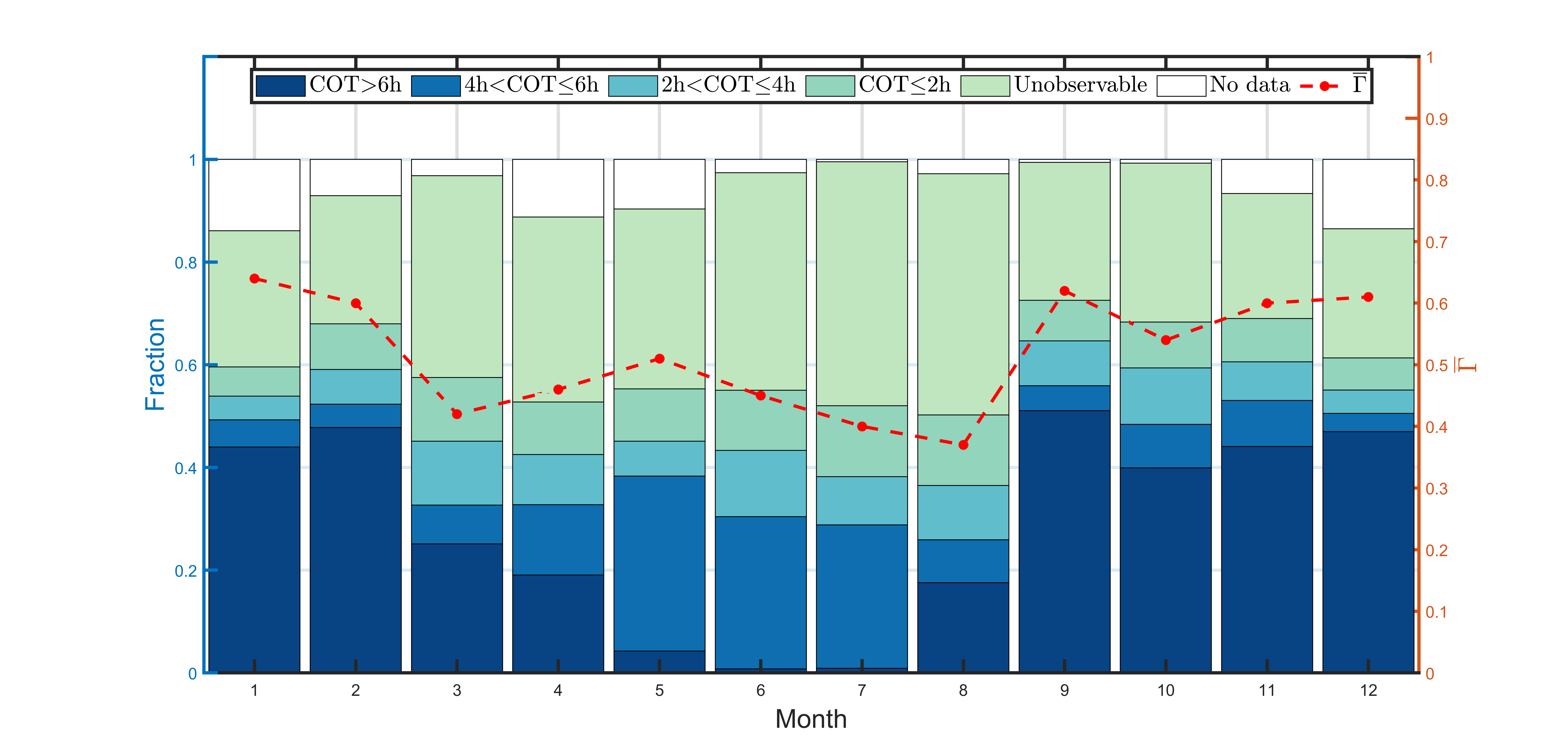}
   \caption{Monthly mean $\Gamma$ value (red dots) and the proportion of different length COT over nighttime at the Muztagh-ata site.}
   \label{fig:f4}
\end{figure*}

\section{Analysis for the Muztagh-ata Site}
The nights for which we have computed $\Gamma$ values all had a record of all-sky images for more than 50\% of the nighttime. Over the five years, we have gathered a total of 1598 nights for $\Gamma$ calculation, constituting 87\% of the entire period. The statistical outcomes for these 1598 nights are presented in Figure~\ref{fig:f3}. There were 522 nights exhibiting $\Gamma\geq0.9$, signifying absolute clear nights, while 888 nights surpassed the threshold of $\Gamma\geq0.36$. This site can provide 122 absolutely clear nights and 205 very good nights annually.

Further, we have conducted monthly statistical analysis to discern the fractions of four ranges of COT within nighttime. The outcomes are depicted in Figure~\ref{fig:f4}, wherein the red line traces the monthly mean $\Gamma$ values. Through this approach, we identify January as the month with the best nighttime cloud coverage, with an average $\Gamma$ value of 0.64. Comprehensive details regarding the mean number of nights characterized by
$\Gamma\geq0.9$, $\Gamma\geq0.4$ and $\Gamma\geq0.36$ are enumerated in Table~\ref{tab:t1}.

\begin{table}[h]
    \centering
      \caption{Monthly statistic of the number of nights $\Gamma\geq0.9$, $\Gamma\geq0.4$ and $\Gamma\geq0.36$ at the Muztagh-ata site}
\label{tab:t1} 
\begin{tabular}{ccc}
\hline
Month                 & $\Gamma\geq0.9$ & $\Gamma\geq0.36$\\
\hline
\hline
Jan	&	14.3&	22	\\
Feb	&	11.9&	18	\\
Mar	&	6.9&15.3	\\
Apr	&	6.5&14.8	\\
May	&	11.2&	17.1	\\
Jun	&	8.1&15.3	\\
Jul	&	7.1&13.6	\\
Aug	&	5.8&12.6	\\
Sep	&	14.7&20	\\
Oct	&	10.5&18.6	\\
Nov	&	11.2&19.9	\\
Dec	&	14.2&19.7	\\
yearly & 122.4&205.4   \\
\hline
\end{tabular}
\end{table}

In our previous paper\citep{2023RAA....23d5015X}, An observing night is defined as one with AOT exceeding 50\%, while a clear night is designated when the AOT surpasses 75\%. The Muztagh-ata site can provide an average of 175 clear nights and 227 observing nights per year. Within the dataset of 1598 nights spanning from 2017 to 2021, we identified 744 out of 812 clear nights where the COT exceeded 50\% of the night, and 902 out of 1052 observing nights with COT extending beyond 40\%. These statistics indicate that the presence of passing clouds is not a frequent disruption during either clear or observing nights at the Muztagh-ata site.The site can offer approximately 160 clear nights with continuous observing time exceeding 50\% and 195 observing nights with continuous observing time exceeding 40\% each year. A summary of the proportions for clear, observing nights and Non-observing nights corresponding to different lengths of COT is presented in Table~\ref{tab:t2}. Additionally, we analyze the fragmentation of observing time across the rest of 546 nights. Results reveal that 126 of these nights still exhibited COT extending beyond 20\%, which translates to approximately 61 hours of potential observing time per year during Non-observing nights.

\begin{table}[htb]
    \centering
    \caption{Proportion of different lengths of COT in clear and observing nights.}
    \label{tab:t2} 
    \begin{threeparttable}
        \begin{tabular}{lccccc}
            \hline\hline
            \multicolumn{1}{c}{ } &  \multicolumn{4}{c}{COT} \\
            \cline{2-6}
             &$\geq 20\%$ & $\geq 30\%$ & $\geq 40\%$ & $\geq 50\%$ & $\geq 60\%$ \\
            \hline
            clear nights ( 812 days) &   $100\%$ & 99\% & 96\% & 92\% & 85\% \\
            observing nights ( 1052 days) &99\% & 94\% & 86\% & 77\% & 69\% \\
            Non-observing nights ( 546 days) &23\% & 10\% & $\sim$ & $\sim$ & $\sim$ \\

            \hline
        \end{tabular}
    \end{threeparttable}
\end{table}

\section{Conclusion}
In this paper, we introduce a fresh approach for quantitatively assessing nighttime cloud coverage at astronomical observatories and potential sites. This method provides a comprehensive evaluation by considering both the extent of AOT and its fragmentation. Employing this method, we analyze the fragmentation of AOT across an all-sky image dataset encompassing 1598 nights, spanning the five years from 2017 to 2021 at the Muztagh-ata site. The outcomes verify the validity of this approach.

Our study yields more accurate monthly statistics regarding nighttime cloud coverage at the Muztagh-ata site.
We ascertain that the site offers approximately 122 absolute clear nights each year, as well as 205 very good nights each year.
Absolute clear nights offer a minimum of 90\% or more observation time, with the fraction for continuous overnight observations exceeding 90\%. very good nights provide an average of over 60\% observation time, with the fraction of continuous overnight observations surpassing 40\%, ensuring at least 50\% of observable time.
Remarkably, over 92\% of clear nights and 77\% of observing nights with COT extending beyond 50\% of the nighttime hours are realized at this site. These findings reveal the remarkable advantages of nighttime sky clarity this site presents for astronomical observations.

\begin{acknowledgements}
This work was supported by the Chinese Academy of Science (CAS) “Light of West China” Program(\text{No.2022\_XBQNXZ\_014}), the Joint Research Fund in Astronomy under a cooperative agreement between the National Natural Science Foundation of China (NSFC), the Chinese Academy of Sciences (CAS) (Grant No.U2031209), the Xinjiang Natural Science Foundation (Grant No.2022D01A357) and the National Natural Science Foundation of China (Grant No.11873081).
\end{acknowledgements}
\bibliographystyle{raa}
\bibliography{reference}  

\end{document}